\documentclass[a4paper,11pt]{article}
\pdfoutput=1 

\usepackage{jheppub} 

\usepackage{slashed}

\title{Fine-tuned Spin-3/2 and the Hierarchy Problem}

\author[a,1]{Ozan Sarg{\i}n,\note{Corresponding author.}}

\affiliation[a]{{\.I}zmir Institute of Technology,\\Department of Physics, TR35430, {\.I}zmir, Turkey} 

\emailAdd{ozansargin@iyte.edu.tr}

\abstract{In the past,  \textit{Kundu et al.}  and \textit{Chakraborty et al.} used extra scalar fields to cancel the quadratic divergences in the Higgs mass squared and they determined the mass of the required scalar field. In this work, a spin-3/2 particle has been used in the same manner to nullify the power-law divergences and it is determined that the mass of the spin-3/2 particle resides in the ball park of the GUT scale.}

\begin{document}
\maketitle
\flushbottom

\section{Introduction}
\label{sec:intro}
 With the discovery of a resonance at the LHC \cite{Gaad,Chatrchyan} that seems to be rather consistent with the standard model  Higgs boson in light of the initial assessment of its properties \cite{Ellis,Djouadi}; the electroweak naturalness \cite{Weisskopf,Wilson} is the foremost problem that we should turn our attention to. Even though many new physics theories have been suggested in order to cure the destabilization of the Higgs mass \cite{Susskind,Weinberg,Dimopoulos,Arkani-Hamed,Randall,Arkani-Hamed2}; so far no signal of these  has been observed \cite{Flechl,Feng3,Rappoccio}. The lack of any new physics particle at the ${\rm TeV}$ scale casts doubts on the relevance of the idea of naturalness and strengthens the view that the Standard Model (SM) of electroweak and strong interactions may well be the model of physics at Fermi scale \cite{Altarelli,Giudcie,Feng,Wells}. However, this situation does not change the fact that SM is an incomplete, effective theory since, for one thing, it is missing the essential dark matter (DM) candidate \cite{Bertone,Feng2,Porter} and for another, it offers no dynamical principle that generates the masses and couplings of the theory \cite{Ahmad,Fukuda,Ahn,Abe}.
\par
 Since  requiring that the SM be technically natural brings us to a dead end in terms of the guidelines to New Physics (NP), we explore  an orthogonal possibility here and assume that the electroweak scale is stabilized via a mechanism based on the fine-tuning of a sector which is split from the SM. The same line of reasoning has been employed before by  \textit{Kundu et al.} \cite{Kundu} and \textit{Chakraborty et al.} \cite{Chakraborty} through different models based on singlet scalars. The main motive behind  this approach is to cancel the power-law divergences in the Higgs mass squared via the loops of extra fields and  fine-tuning the parameters of  the specific BSM model that is utilised \cite{Bazzocchi,Andrianov}. Even though this is intrinsically a fine-tuning operation, it brings along valuable advantages such as accommodating viable dark matter candidates \cite{McDonald,Demir2,Guo}. However, exploiting real scalar fields for this, comes with its own drawbacks. Since real singlet scalars with a vacuum expectation value are bound to mixing with the CP-even component of the SM Higgs field itself; an all-encompassing, simultaneous cancellation is not achievable \cite{Karahan}.
 \par
 In the present work, we study the Higgs mass stabilization problem by a  hidden spin-3/2 particle high above the electroweak scale and examine the radiative corrections it induces on the Higgs self energy in an effective field theory approach using cut-off regularization so as to obtain an estimate of the mass of this new particle by demanding that the total one loop corrections to the Higgs mass should cancel. The main advantage of our model over the singlet scalar approaches is that while the latter need auxiliary fields such as vector-like fermions in order to stabilize the NP sector itself, hidden spin-3/2 field is free from such requirements. Due to the unique character of our spin-3/2 interaction with the SM, it is impossible to observe a spin-3/2 particle on mass shell. This means that the BSM sector in our model is a genuinely stable hidden extension of SM. This constitutes a phenomenological advantage, which has important implications not just for the electroweak stabilization but also for the Higgs boson and  hidden sector correlation.
 Another implication is related to the fact that  our calculations reveal that this higher spin particle resides in the ball park of GUT scale. If ongoing searches at the LHC reveal no particles  at the ${\rm TeV}$ scale combined with  the fact that the next higher spin particle (spin-3/2) inhabits the GUT scale may strengthen the grand desert notion in the GUTs without ${\rm TeV}$ scale NP \cite{Dimopoulos2}.

\section{The Hierarchy Problem}
\label{sec:hierarchy}
 Spontaneous breaking of electroweak symmetry  is implemented in the SM by postulating the existence of a fundamental scalar, the Higgs field, whose potential is parameterized by a dimensionful
 mass-squared parameter $\mu^{2}$ and a dimensionless Higgs self-coupling $\lambda$. The Higgs field takes on a constant value everywhere in space-time called the vacuum expectation
 value (\emph{vev}) $v = \sqrt{\mu^{2}/\lambda}= 246\ {\rm GeV}$, which being a dimensionful parameter sets the scale for all the masses of the theory in terms of the Yukawa couplings.

\par
 Fundamental scalars are widely considered as unnatural. As opposed to the fermions and gauge bosons whose masses are under control by chiral and gauge symmetries respectively, masses of fundamental scalars are not protected by any kind of symmetry. This makes them vulnerable to divergent radiative corrections  they get from loop diagrams. The technical usage of the term naturalness is related to the quantum corrections that a parameter gets when one makes use of perturbation theory to calculate the properties of a theory.
\par
 The hierarchy problem we consider here is specifically associated with the quantum corrections to the Higgs mass-squared $m_{h}^{2}$. A simple statement of the problem can be given as follows. In an effective field theory with a hard ultraviolet cut-off $\Lambda$, loop diagrams induce  radiative corrections in the Higgs self energy  such that
\begin{equation}\label{e1}
  m_{h}^{2}=(m_{h}^{2})_{bare}+\mathcal{F}(\lambda , g_{i}^{2})\Lambda^{2},
\end{equation}
 where $m_{h}=\sqrt{2\lambda}v$ is the physical Higgs boson mass, $g_{i}$ are the renormalized gauge couplings of the SM and the second term signifies the $\mathcal{O}(\Lambda^{2})$ quantum corrections.
 If the second term in equation (\ref{e1}) is of the same order or smaller than the measured value of the Higgs mass, it is said that the parameter is natural; however, if  the measured value turns out to be  much smaller than the radiative correction term, it is said that the theory is unnatural because this hints at a contingent cancellation  between the bare mass and the quantum effects  so as to produce the measured value of the Higgs boson mass.
\par
    Originally put forward by Veltman \cite{Veltman}, the SM one-loop corrections to the Higgs boson mass reads
\begin{equation}\label{e2}
  (\delta m_{h}^{2})_{SM}=\frac{\Lambda^{2}}{8 \pi^{2}}\Big(-6\lambda_{t}^{2}+\frac{9}{4}g^{2}+\frac{3}{4}g'\,^{2}+6\lambda\Big)
\end{equation}
 where $g$  and $g'$ are the $SU(2)_{L}$ and $U(1)_{Y}$ gauge couplings in the SM respectively, and $\lambda_{t}$ is the top quark Yukawa coupling. Only the top quark Yukawa coupling appears in equation (\ref{e2}) because the contributions of  other fermions  are considerably small. Veltman stated that the Higgs mass should be stable against loop corrections and  the above criterion is used as a means to estimate the Higgs boson mass, hence this expression is commonly called the Veltman condition (VC).
\par
 The Higgs mass estimated using VC is in conflict with the experimental value today and we are in a bit of quandary. Considering the fact that the cutoff regulator can get as high as the Planck mass, we are faced with an unnaturalness of 32 orders of magnitude. If we require that the Higgs  be technically natural, there should appear new physics  around ${\rm TeV}$ scale and remove the quadratic dependence on the cutoff scale $\Lambda$.
\par
 People following this motivation have come up with many NP theories and chief among them is Supersymmetry \cite{Lykken,Martin}. However, despite all the extensive searches, no compelling evidence in favor of any of these NP theories has been observed during the LHC runs reaching  well above the ${\rm TeV}$ scale.
\par
 Having no ${\rm TeV}$ scale NP to prevent the destabilization of the Higgs boson, new mechanisms have been put forward which involve extensions beyond the SM and general relativity. One such mechanism which makes use of conformal symmetry has been introduced by W. A. Bardeen in 1995 \cite{Bardeen}, and paved the way for many variants since then. Anti-gravity effects have also been claimed to be viable  in improving the naturalness of electroweak sector \cite{Salvio}. Another interesting possibility is exploiting the coupling between the Higgs boson and the space-time curvature as a means of harmless, soft fine-tuning that was shown to be capable of solving the hierarchy problem \cite{Demir3}.
  \par
 Aside from all these different approaches, a general practice is to invoke a cancellation mechanism which involves fine-tuning of counter terms that mixes together both low and high scale physical degrees of freedom. The scheme that has been followed here in this work is in line with the aforementioned cancellation mechanism approach but it is a completely new model with a unique interaction.
 The next section is devoted to the description of this model. We first start by a quick review of spin-3/2 and then continue on with the details of the interaction Lagrangian. The interaction is such that, at the renormalizable level it is only through the neutrino portal that spin-3/2 makes contact with the SM. Due to the special constraints that these fields should obey, they can  participate interactions only as virtual particles. Without further ado, lets get to the details in the next section.

\section{The Model}
\label{sec:model}
Spin-3/2 fields, commonly called vector spinors, $\psi_{\mu}$, are introduced to the literature for the first time  by
Rarita and Schwinger \cite{Rarita:1941mf}. The propagator for the $\psi_{\mu}$ reads
\begin{eqnarray}
S^{\alpha\beta}(p) = \frac{i}{{\slashed{p}} - M} \Pi^{\alpha\beta}(p),
\end{eqnarray}
and it involves one spin-3/2 and two spin-1/2 components embedded in the
projector \cite{pilling}
\begin{eqnarray}
\label{project}
\Pi^{\alpha\beta} = -\eta^{\alpha\beta} +
\frac{\gamma^{\alpha}\gamma^{\beta}}{3}+
\frac{\left(\gamma^{\alpha}p^{\beta} -
\gamma^{\beta}p^{\alpha}\right)}{3M}+\frac{2
p^{\alpha}p^{\beta}}{3 M^2},
\end{eqnarray}
that exhibits both spinor and vector characteristics.
In order to remove the two spin-1/2 components we impose the two constraints \cite{demir,Pascalutsa:2000kd}
\begin{eqnarray}
\label{eqn4}
p^{\mu}\psi_{\mu}(p)\rfloor_{p^2=M^2}=0,
\end{eqnarray}
and
\begin{eqnarray}
\label{eqn4p}
\gamma^{\mu}\psi_{\mu}(p)\rfloor_{p^2=M^2}=0,
\end{eqnarray}
 after which $\psi_{\mu}$
satisfies  the Dirac equation
\begin{eqnarray}\label{eqn5}
\left(\slashed{p} - M\right)\psi_{\mu}=0
\end{eqnarray}
as expected of an on-shell fermion. The constraints (\ref{eqn4}) and (\ref{eqn4p}) imply that $p^{\mu}\psi_{\mu}(p)$ and $\gamma^{\mu}\psi_{\mu}(p)$ both vanish on the physical shell $p^2=M^2$.

Characteristic of singlet fermions, the $\psi_{\mu}$, at the renormalizable level, makes contact with the SM via
\begin{eqnarray}
\label{int1}
{\mathcal{L}}^{(int)}_{3/2} = c^{i}_{{3/2}} \overline{L^{i}} H \gamma^{\mu}\psi_{\mu} + h.c.
\end{eqnarray}
in which
\begin{eqnarray}
L^i = \left(\begin{array}{c}\nu_{\ell L}\\ \ell_L\end{array}\right)_{i}
\end{eqnarray}
is the lepton doublet ($i=1,2,3$), and
\begin{eqnarray}
H = \frac{1}{\sqrt{2}}\left(\begin{array}{c}v + h + i \varphi^0\\ \sqrt{2} \varphi^{-}\end{array}\right)
\end{eqnarray}
is the Higgs doublet with vacuum expectation value $v\approx 246\ {\rm GeV}$, Higgs boson $h$, and Goldstone bosons  $\varphi^{-}$, $\varphi^0$ and $\varphi^+$ (forming the longitudinal components of $W^{-}$, $Z$ and $W^{+}$ bosons, respectively).
\par
\begin{figure}[ht!]
  \begin{center}
  \includegraphics[scale=0.65]{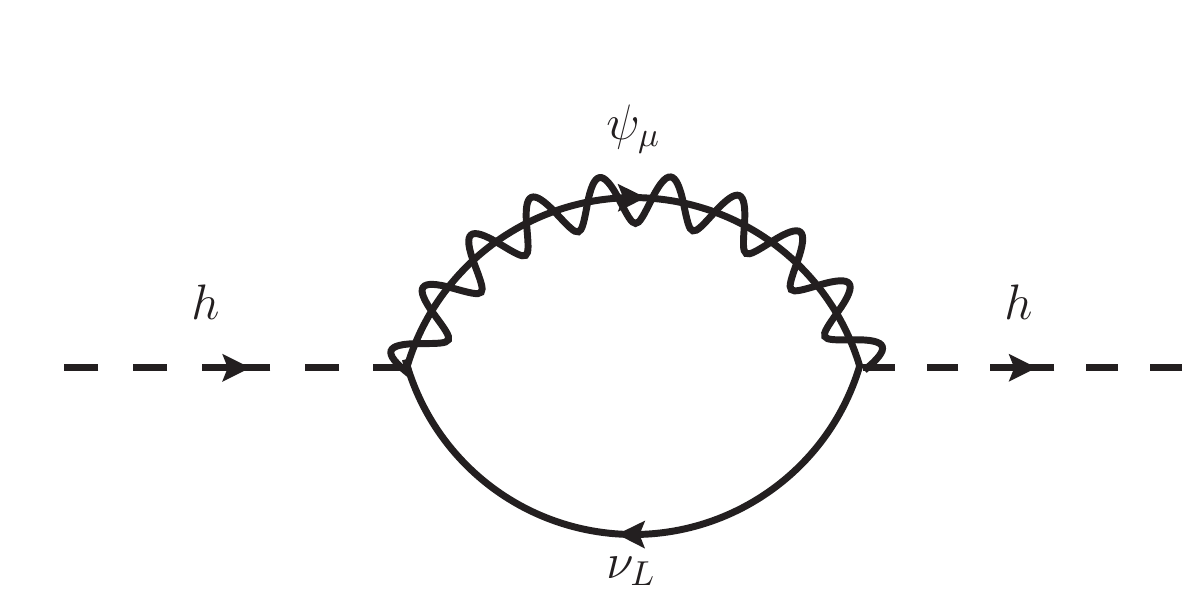}
  \end{center}
  \caption{The $\psi_{\mu}-\nu_L$ loop that generates the quartic correction in the Higgs mass.} \label{fig1} 
\end{figure}

Because of the constraint in equation (\ref{eqn4p}),  effects of the vector spinor is mainly restricted to the loop diagrams. One such diagram is depicted in the Fig. \ref{fig1}. As a result of this interaction, $\psi_{\mu}$ contributes to the Higgs boson mass correction given in equation (\ref{e2}).

\par
The contribution of the  Feynman diagram in the Fig. \ref{fig1} is given by

\begin{eqnarray}\label{diagx}
  -i\Sigma \delta^{4}(p-p^{'})(2\pi)^{4}=\qquad\qquad\qquad\qquad\qquad\qquad\qquad\qquad\qquad\qquad     \\ \nonumber
  -Tr\Bigg\{\int\frac{d^{4}q_{1}}{(2\pi)^{4}}\frac{d^{4}q_{2}}{(2\pi)^{4}}
  \Big(\frac{i c_{3/2}}{\sqrt{2}}\gamma_{\alpha}P_{L}\Big)\Bigg[\;\,\frac{i(\slashed{q_{1}}+M)}{q_{1}^{2}-M^{2}}\;\,\Pi^{\alpha\beta}\;\,\Bigg]
  \Big(\frac{i c_{3/2}}{\sqrt{2}}\gamma_{\beta}P_{L}\Big)\Big(\frac{i \slashed{q_{2}}}{q_{2}^{2}}\Big)\\ \nonumber
  (2\pi)^{4}\delta^{4}(p+q_{2}-q_{1})(2\pi)^{4}\delta^{4}(q_{1}-q_{2}-p^{'})\Bigg\}.
\end{eqnarray}

\par
The term in the square brackets is the propagator of $\psi_{\mu}$ where $\Pi^{\alpha\beta}$ is the projector  as a function of the loop momentum $q_{1}$ of $\psi_{\mu}$, the explicit form of which is

\begin{eqnarray}\label{projq1}
\Pi^{\alpha\beta} = -\eta^{\alpha\beta} +
\frac{\gamma^{\alpha}\gamma^{\beta}}{3}+
\frac{\left(\gamma^{\alpha}q_{1}^{\beta} -
\gamma^{\beta}q_{1}^{\alpha}\right)}{3M}+\frac{2
q_{1}^{\alpha}q_{1}^{\beta}}{3 M^2}\;.
\end{eqnarray}

The leftmost vertex is designated by $\alpha$ and the other by $\beta$, hence the two vertex factors are

\begin{equation}\label{vert1}
 i\frac{c_{{3/2}}}{\sqrt{2}}\gamma_{\alpha}P_{L}
\end{equation}
and
\begin{equation}\label{vert2}
 i\frac{c_{{3/2}}}{\sqrt{2}}\gamma_{\beta}P_{L}
\end{equation}
 respectively.
 \par
 After taking the integral over the loop momentum of the neutrino $q_{2}$, the expression takes the form
\begin{eqnarray}\label{diagx2}
  i\Sigma=\frac{c_{{3/2}}^{2}}{2}\int\frac{d^{4}q_{1}}{(2\pi)^{4}}Tr\Bigg\{\gamma_{\alpha}P_{L}\;\,\frac{(\slashed{q_{1}}+M)}{q_{1}^{2}-M^{2}}\;\,
  \Big[-\eta^{\alpha\beta} +
\frac{\gamma^{\alpha}\gamma^{\beta}}{3}+ \\ \nonumber
\frac{\left(\gamma^{\alpha}q_{1}^{\beta} -
\gamma^{\beta}q_{1}^{\alpha}\right)}{3M}+\frac{2
q_{1}^{\alpha}q_{1}^{\beta}}{3 M^2}\Big]
  \gamma_{\beta}P_{L}\;\,\frac{\slashed{q_{1}}} {q_{1}^{2}}\Bigg\}\; .
\end{eqnarray}
\par
There are five traces that should be evaluated and they are denoted by $a$, $b$, $c$, $d$ and $e$ in the following:
\begin{eqnarray}\label{tracea}
  a\equiv Tr\Bigg\{\gamma_{\alpha}P_{L}(\slashed{q_{1}}+M)
  \eta^{\alpha\beta}
  \gamma_{\beta}P_{L}\slashed{q_{1}}\Bigg\}=-4 q_{1}\cdot q_{1} 
\end{eqnarray}

\begin{eqnarray}\label{traceb}
  b\equiv Tr\Bigg\{\gamma_{\alpha}P_{L}(\slashed{q_{1}}+M)
  \gamma_{\beta}P_{L}\slashed{q_{1}}\Bigg\}= 4 q_{1\, \alpha}q_{1\, \beta}-2\eta_{\alpha\beta}\, q_{1}\cdot q_{1}
\end{eqnarray}

\begin{eqnarray}\label{tracec}
  c\equiv Tr\Bigg\{\gamma_{\alpha}P_{L}(\slashed{q_{1}}+M)
  \gamma^{\alpha}\gamma^{\beta}\gamma_{\beta}P_{L}\slashed{q_{1}}\Bigg\}= -16 q_{1}\cdot q_{1} 
\end{eqnarray}

\begin{eqnarray}\label{traced}
  d\equiv Tr\Bigg\{\gamma_{\alpha}P_{L}(\slashed{q_{1}}+M)
  \gamma^{\alpha}\gamma_{\beta}P_{L}\slashed{q_{1}}\Bigg\}= 8 M q_{1\, \beta} 
\end{eqnarray}

\begin{eqnarray}\label{tracee}
  e\equiv Tr\Bigg\{\gamma_{\alpha}P_{L}(\slashed{q_{1}}+M)
  \gamma^{\beta}\gamma_{\beta}P_{L}\slashed{q_{1}}\Bigg\}= 8 M q_{1\, \alpha}\;. 
\end{eqnarray}

\par
Equation (\ref{diagx2}) in terms of the five traces can be written as

\begin{eqnarray}\label{diagx3}
  \Sigma=\frac{i c_{{3/2}}^{2}}{2}\int\frac{d^{4}q_{1}}{(2\pi)^{4}}\frac{\Big[a-\frac{2
q_{1}^{\alpha}q_{1}^{\beta}}{3 M^2}b-\frac{c}{3}-\frac{\left(q_{1}^{\beta}d -
q_{1}^{\alpha}e\right)}{3M}\Big]}{(q_{1}^{2}-M^{2})q_{1}^{2}}
\end{eqnarray}
\par
After replacing the values of the traces, the term in the square brackets in equation (\ref{diagx3}) becomes

\begin{eqnarray}\label{nomsqu}
  \Big[a-\frac{2
q_{1}^{\alpha}q_{1}^{\beta}}{3 M^2}b-\frac{c}{3}-\frac{\left(q_{1}^{\beta}d -
q_{1}^{\alpha}e\right)}{3M}\Big]=\frac{4}{3}\Big(q_{1}\cdot q_{1}- \frac{(q_{1}\cdot q_{1})^{2}}{M^2}\Big). 
\end{eqnarray}
Plugging equation (\ref{nomsqu}) in equation (\ref{diagx3}), it is possible to split the integral over $q_{1}$ into two halves
\begin{eqnarray}\label{diagx4}
  \Sigma=\frac{2i c_{{3/2}}^{2}}{3}\Bigg\{\int\frac{d^{4}q_{1}}{(2\pi)^{4}}\frac{1}{(q_{1}^{2}-M^{2})}-\frac{1}{M^{2}}\int\frac{d^{4}q_{1}}{(2\pi)^{4}}\frac{q_{1}^{2}}{(q_{1}^{2}-M^{2})}\Bigg\}\;.
\end{eqnarray}
Denoting the fist integral by $I_{1}$ and the second by $I_{2}$, equation (\ref{diagx4}) takes the form

\begin{eqnarray}\label{diagx5}
  \Sigma=\frac{2i c_{{3/2}}^{2}}{3}\Bigg\{I_{1}-\frac{1}{M^{2}}I_{2}\Bigg\}\;.
\end{eqnarray}
Using cutoff regularization, the two integrals $I_{1}$ and $I_{2}$ are evaluated to be
\begin{eqnarray}\label{integ1}
  I_{1}=-\frac{i}{16\pi^{2}}\Bigg[\Lambda^{2}-M^{2}\ln\Big(\frac{\Lambda^{2}+M^{2}}{M^{2}}\Big)\Bigg]  ,      
\end{eqnarray}

\begin{eqnarray}\label{integ2}
  I_{2}=\frac{i}{16\pi^{2}}\Bigg[\frac{(\Lambda^{2}+M^{2})^2-M^{4}}{2}-2M^{2}\Lambda^{2}+M^{4}\ln\Big(\frac{\Lambda^{2}+M^{2}}{M^{2}}\Big)\Bigg] \; ;      
\end{eqnarray}
where $\Lambda$ designates the cutoff scale.
\par
After inserting equations (\ref{integ1}) and (\ref{integ2}) into (\ref{diagx5}), the mass correction to the Higgs mass due to spin-3/2 particle becomes
\begin{eqnarray}\label{masscor}
  \Sigma=\frac{ c_{{3/2}}^{2}}{48\pi^2}\frac{\Lambda^{4}}{M^{2}}=(\delta m_{h}^{2})_{3/2}\;.
\end{eqnarray}
\par
It is interesting to note that the  mass correction in (\ref{masscor}) is of positive sign and purely quartic (no quadratic correction arises due to spin-3/2) although one would expect it to have the opposite sign since this is a fermion that we are dealing with. The crucial point is that the origin of this quartic contribution can be traced back to the last term in the propagator of the spin-3/2, to wit, the $p^{\alpha}p^{\beta}$ term in equation (\ref{project}) and that tells us that the longitudinal component of the propagator overrides the fermionic character at high energy.
\par
After this remark, now let us get back to the calculation. Recall that the SM one loop correction to the Higgs mass reads
\begin{equation}
  (\delta m_{h}^{2})_{SM}=\frac{\Lambda^{2}}{8 \pi^{2}}\Big(-6\lambda_{t}^{2}+\frac{9}{4}g^{2}+\frac{3}{4}g'\,^{2}+6\lambda\Big)\;.
\end{equation}

\par
The total correction to the Higgs mass is the sum of SM part plus the spin-3/2 contribution; to wit
\begin{equation}
  \delta m_{h}^{2}=(\delta m_{h}^{2})_{SM}+(\delta m_{h}^{2})_{3/2}\;.
\end{equation}
\par
\par
If we allow the possibility that the hidden sector is finely tuned  such that the total quantum correction to $m_{h}^{2}$ vanishes, i.e.
\begin{equation}
  \delta m_{h}^{2}=0,
\end{equation}
we have a clear constraint on the mass scale of the quanta of the spin-3/2 field
\begin{equation}
 \frac{\Lambda^{2}}{8 \pi^{2}}\Big(-6\lambda_{t}^{2}+\frac{9}{4}g^{2}+\frac{3}{4}g'\,^{2}+6\lambda\Big)+\frac{ c_{{3/2}}^{2}}{48\pi^2}\frac{\Lambda^{4}}{M^{2}}=0.
\end{equation}
\par
If we take the cutoff scale to be the Planck mass and the coupling constant of the spin-3/2, $c_{{3/2}}$, of order unity, this gives us  
\begin{equation}
 M\approx 10^{16}\, \mathrm{GeV}
\end{equation}
as the mass scale of spin-3/2 quanta.

\section{Summary}
\label{sec:summary}
In this work we have taken up the Hierarchy problem from a different angle. Contrary to the general acceptance
that electroweak scale should be technically natural, we have allowed the possibility that fine tuning may well be the option that nature favors. The rationale behind this choice is the non-existence of any experimental proof of the natural theories of BSM that predict ${\rm TeV}$-scale NP so as to stabilize the Higgs boson.
\par
We have assumed that the Higgs boson stays stable via a finely tuned hidden sector which involves a spin-3/2 field that is split from the SM and whose sole contact with it at the renormalizable level is through the neutrino portal. The interaction lagrangian of our model is given by equation (\ref{int1}) along with the two constraint equations (\ref{eqn4}) and (\ref{eqn4p}) that these fields should obey so as to satisfy the Dirac equation. The distinctive feature of this model is that the spin-3/2 field is enforced to be inherently off-shell  due to the constraint equation (\ref{eqn4p}). As such, this field is hidden from the SM and  the effects of it is mainly expected to be visible through the loop diagrams.
\par One such diagram that one can witness the effects of $\psi_{\mu}$ is the loop given in Fig. \ref{fig1} which designates the contribution of the spin-3/2 to the Higgs self energy. As such, this diagram plays an important role in the Hierarchy problem.
\par The loop diagram that we have depicted in Fig. \ref{fig1} induces a quartic radiative correction in the Higgs mass. However, contrary to the general expectation that fermions should have negative radiative powerlaw corrections to the Higgs mass, what we observe here is that the spin-3/2 field has a positive contribution. This strange phenomenon can be traced back to the longitudinal terms in the propagator of $\psi_{\mu}$ and reveals that the fermionic character is washed out at high energy.
\par  After making the pivotal assumption that the Higgs mass stays stable via a mechanism that involves a finely tuned hidden sector, we have used the total mass correction to the Higgs mass as a constraint to calculate the mass scale of the spin-3/2 field. This calculation has revealed that the spin-3/2 field is indeed split from the SM because it resides well above the SM with a mass around $M\approx 10^{16}\, \mathrm{GeV}$ which is in the ball park of GUT scale. This finding is a strong support for the  notion called  the grand desert in the GUTs without $\mathrm{TeV}$ scale NP.

\end{document}